\documentclass[11pt,twoside]{article}
\usepackage{asp2010}
\usepackage{natbib}

\aspvolume{471} 
\aspvoltitle{Origins of the Expanding Universe: 1912-1932}
\aspcpryear{2013} 
\aspvolauthor{Michael J. Way and Deidre Hunter, eds.} 

\begin{document}

\title{What Else Did V.~M. Slipher Do?}

\author{Joseph S. Tenn\affil{Department of Physics \& Astronomy,
Sonoma State University, Rohnert Park, CA, 94928, USA}}

\begin{abstract}
When V.~M. Slipher gave the 1933 George Darwin lecture to the 
Royal Astronomical Society, it was natural that he spoke on spectrographic 
studies of planets. Less than one--sixth of his published work deals with 
globular clusters and the objects we now call galaxies. In his most productive 
years, when he had Percival Lowell to give him direction, Slipher made major 
discoveries regarding stars, galactic nebulae, and solar system objects. These 
included the first spectroscopic measurement of the rotation period of Uranus, 
evidence that Venus's rotation is very slow, the existence of reflection nebulae 
and hence interstellar dust, and the stationary lines that prove the existence of 
interstellar calcium and sodium. After Lowell's death in 1916 Slipher 
continued making spectroscopic observations of planets, comets, and the 
aurora and night sky. He directed the Lowell Observatory from 1916 to 1954, 
where his greatest achievements were keeping the observatory running despite 
very limited staff and budget, and initiating and supervising the ``successful" 
search for Lowell's Planet X. However, he did little science in his last 
decades, spending most of his time and energy on business endeavors.
\end{abstract}

\section{Introduction}
Vesto Melvin Slipher, always referred to and addressed as 
``V.~M." \citep{Giclas2007,Hoyt1980BMNAS..52..410H}
came to Flagstaff in August 1901, two months after completing his B.A. in 
mechanics and astronomy at Indiana University, because his professor, Wilbur 
Cogshall, had persuaded Percival Lowell
to hire him temporarily. 
He arrived at age 25 and stayed there 53 years. After retirement he lived 15 
more years in Flagstaff. I will discuss his life in Flagstaff, which I divide into 
five parts, and his research on the night sky, the aurora, planets, comets, stellar 
radial velocities, variable stars, and interstellar gas and dust. We have heard 
from others about his early life and his work on globular clusters and galaxies. 
The closest thing to a published biography is William G. Hoyt's \emph{Biographical 
Memoir} \citep{Hoyt1980BMNAS..52..410H}.

\begin{figure}
\center{\includegraphics[scale=.7]{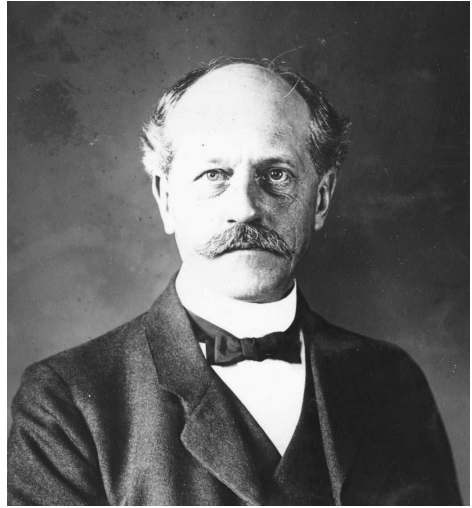}}
\caption{Percival Lowell, 1904 (courtesy Lowell Observatory Archives (LOA))}\label{tennfig01}
\end{figure}

\section{Great Achievements under Lowell, 1901--1916}
In his early years at Lowell Observatory, Slipher was not his own man. This 
was fortunate, as together he and his employer made a formidable team. As 
John S. Hall wrote in an obituary in \emph{Sky and Telescope}
\citep{Hall1970ST....39...84H},

\begin{quote}
Slipher and Lowell had complementary temperaments. The latter 
was brilliant, enthusiastic, and a driving personality. ... Slipher, on 
the other hand, was deliberate, fastidious, patient, and showed a high 
order of technical knowledge.
\end{quote}

   Lowell knew what he wanted, and Slipher provided it. 

   Lowell, of course, was primarily interested in the solar system, with 
special emphasis on Mars. He wanted Slipher to find chlorophyll, as well as 
oxygen and water there. As we have heard, he asked Slipher to obtain a 
spectrogram of a spiral nebula because he thought of it as a newly-forming 
solar system. However, he allowed Slipher to spend some of his time on his 
own pursuits, and Slipher was interested in the then-fashionable fields of 
measuring stellar radial velocities and discovering spectroscopic binary stars. 
His first publication, in the \emph{Astronomical Journal} in 1902 
\citep{Lowell1902AJ.....22..190L}, was a report of measurements of the variable 
velocity of zeta Herculis with the new spectrograph.

   In the paper he compares his average radial velocity of --74.4 with the --74.6
reported earlier by Lick Observatory's W.~W. Campbell, who was swiftly 
becoming recognized as the world's leading astronomical spectroscopist. (As 
a teacher I would take points off for not specifying units. This is especially 
bad because Slipher sometimes discussed velocities in miles per second. I had 
to go to Campbell's article \citep{Campbell1902ApJ....16..114C} to find 
that the velocities were in ``km", at the time the standard abbreviation for 
km/s. It is possible that young Slipher was unaware of this convention.)

   Volume 1 of the \emph{Lowell Observatory Bulletin}, dating from 1903 to 1911, 
shows that Slipher was already an important member of the small Lowell 
team. The volume contains 62 articles, 13 of them by Slipher alone and one 
co-authored by him. There are 38 by Lowell, some of them including 
spectroscopic observations by Slipher, and 10 by other members of the staff. 

Slipher's publications in this early volume present spectroscopic 
observations of stars, including spectroscopic binary stars, standard velocity 
stars, and stars of variable radial velocity, of the Moon and planets, and of 
Halley's comet. He also began an extensive study of the Crab Nebula, which 
he never published. 

   Among his most important discoveries in this period were two involving 
the interstellar medium. In 1909 he published an account
\citep{Slipher1909LowOB...2....1S} of the selective absorption of
light in space, 
proof that there were calcium ions in the interstellar medium between the Sun 
and a number of stars in Scorpius, Orion, Ophiuchus, and Perseus. In each 
case the sharp, weak calcium lines remained stationary while lines from the 
binary stars shifted back and forth. This confirmed a hypothesis made earlier 
by Johannes Hartmann of Potsdam, who found stationary calcium and sodium 
lines in Nova Persei in 1901 \citep{Vogel1901ApJ....13..217V} and 
calcium again in the single-line spectroscopic binary delta Orionis in 1904 
\citep{Hartmann1904ApJ....19..268H}.  According to historian Daniel 
Seeley \citep[][p. 83]{Seeley1973PhDT........12S}, ``Hartmann set the stage 
for investigations into interstellar gas but Slipher provided the first real 
progress -- his observations indicated that the interstellar lines were not a 
singular phenomenon and his interpretation proved to be accurate."
However, Seeley also notes, ``Slipher's interpretations of the stationary line 
data, published in a \emph{Lowell Observatory Bulletin}, either were not widely 
known or were ignored."\footnote{\citet[][p. 84]{Seeley1973PhDT........12S}}

   Slipher found the first reflection nebula, evidence of what we now call 
interstellar dust, in 1912 \citep{Slipher1912LowOB...2...26S}. He noted 
that he found the spectrum of the cloud surrounding Merope, a star in the 
Pleiades, to be identical to that of the star, and that this could be explained by 
assuming that ``the nebula is disintegrated matter similar to what we know in 
the solar system, in the rings of Saturn, comets, etc., and ... it shines by 
reflected star light." However, he ended this paper, published in December 
1912 while he was in the midst of obtaining his measurement of the huge 
velocity of approach of the Andromeda Nebula, with

\begin{quote}
The observation of the nebula in the Pleiades has suggested to me 
that the Andromeda Nebula and similar spiral nebulae might consist 
of a central star enveloped and beclouded by fragmentary and 
disintegrated matter which shines by light supplied by the central 
sun. This conception is in keeping with spectrograms of the 
Andromeda Nebula made here and with Bohlin's value for its 
parallax.
\end{quote}

\begin{figure}
\center{\includegraphics[scale=0.9]{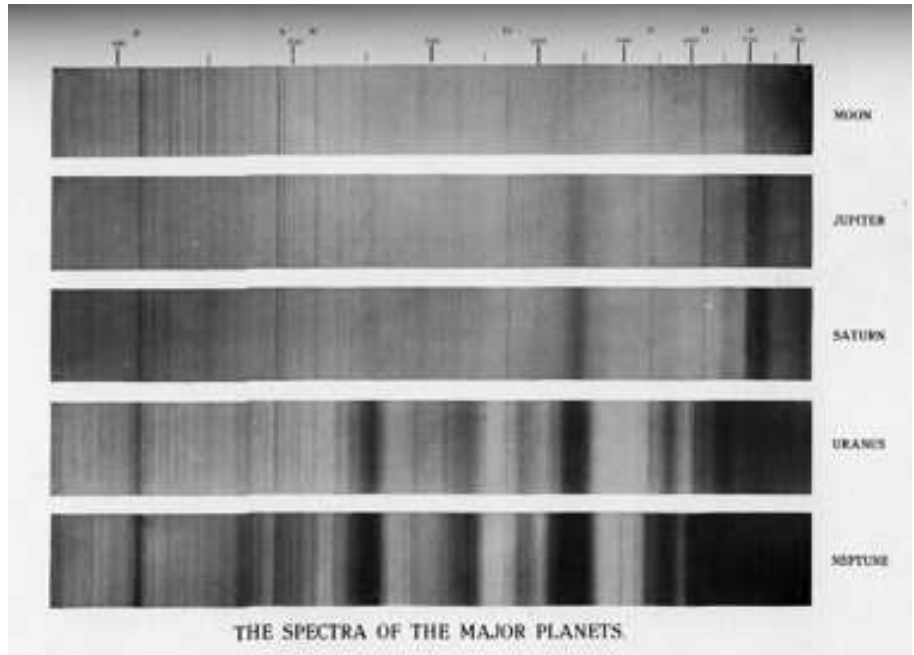}}
\caption{The Spectra of the Major Planets \citep{Slipher1909LowOB...1..231S}.}\label{tennfig02}
\end{figure}

   Also of considerable importance -- it was cited on the awarding of two of 
his gold medals in the 1930s
\citep{Stratton1933MNRAS..93..476S,Einarsson1935PASP...47....5E} -- was his work
on the planets. As early as 1903 he showed that his spectrograph could measure
the rotation period of Mars \citep{Slipher1903LowOB...1...19S}.
He obtained a period 
of 25 h 35 min, ``or just one hour longer than the true period." At a time when 
many thought the rotation period of Venus was about 24 hours he showed that 
it had to be far longer than that. In fact the rotation was too slow to measure. The 
following year he published spectrograms of Uranus and Neptune, and 
compared them with the purely solar radiation from the Moon. By 1906 he 
had added Jupiter and Saturn. After experimenting with new sensitizing dyes 
on his plates, he found a combination which allowed him to be the first to 
extend his spectrograms past 7000 {\AA}ngstroms into the red, so in 1909 he 
provided completely new analyses of the spectra of all four major planets 
(Figure \ref{tennfig02}). He found new absorption bands in these planets, stronger in the 
more distant ones, and he failed to find evidence of oxygen in any of them.

   He was able to show that his spectrograph could detect the rotation of 
Uranus \citep{Slipher1912LowOB...2...19S}. He had tried in 1903 
without success. Six years later Lowell pointed out to him that the line of sight 
component of the rotational motion had increased, and he tried again. By 1911 
he obtained seven good spectrograms, and his results were published the 
following year. Both he and Lowell measured the plates, without knowing 
their orientation. His final result was a rotation period of 10h 50 min, not 
particularly good by today's standards (the current accepted value is 17h 
14min), but the first to be measured.

   It was during this period that Wilbur Cogshall, who had taught V.~M. 
astronomy at Indiana University, wrote him \citep{Cogshall1908a} and
suggested that the University might award him a Ph.D. for 
research he had done at Lowell. V.~M. was enthusiastic, replying, ``Your letter 
was received a few days its content surprised me for the P.H.D. degree has 
been furthest from my thoughts. ... I hardly feel deserving of the honor, ...."  
\citep{Slipher1908}. He sent what he considered ``by far, my best 
work" -- a published paper on the spectra of the planets -- to serve as his 
thesis, but almost lost the degree when Lowell declined to allow him to go to 
Bloomington in June 1908 to defend his thesis. Cogshall suggested
\citep{Cogshall1908b} that Lowell was offended that Slipher was even asked to
defend. The patrician Bostonian thought that the degree should simply be
conferred.  Slipher received the degree a year later, with all residence and
course requirements waived.

    By the time of Lowell's death the day after Slipher's 41st birthday, V.~M. 
had begun examining the night sky with heroic exposures
\citep{Slipher1916LowOB...3....1S}, discovering what he called the permanent 
aurora, with a greenish line known from aurorae present in all his 
spectrograms. He had also added observations of nebulae and interpretations 
involving the interstellar medium to his published work. And of course he had 
observed what we now call galaxies, but this gathering has heard plenty about 
that.

\section{Lowell's Death Brings New Responsibilities and Cares, 1916--1926}

\begin{figure}
\center{\includegraphics[scale=0.6]{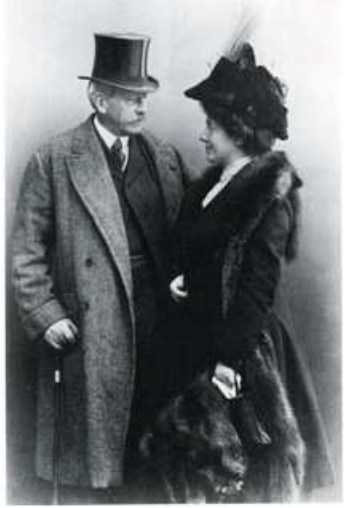}}
\caption{Lowell wedding, 1908 (courtesy LOA).}\label{tennfig03}
\end{figure}

Lowell's unexpected death on 12 November 1916, just eight years after his 
marriage (Figure \ref{tennfig03}) and one year after appointing V.~M. assistant director
and designated successor, was a disaster for his observatory and its staff. For the 
next decade successive trustees fought legal battles with Lowell's widow over 
the estate. She received half the income, and it was a struggle to keep the 
observatory open. As acting director, V.~M. had to be extremely parsimonious. 
Slipher had married in 1904 and his children were nine and five when Lowell 
died. He was justifiably concerned with supporting his family. The daily 
ration of milk from the Observatory cow, Venus, was helpful but not 
sufficient.

\begin{figure}
\center{\includegraphics[scale=0.8]{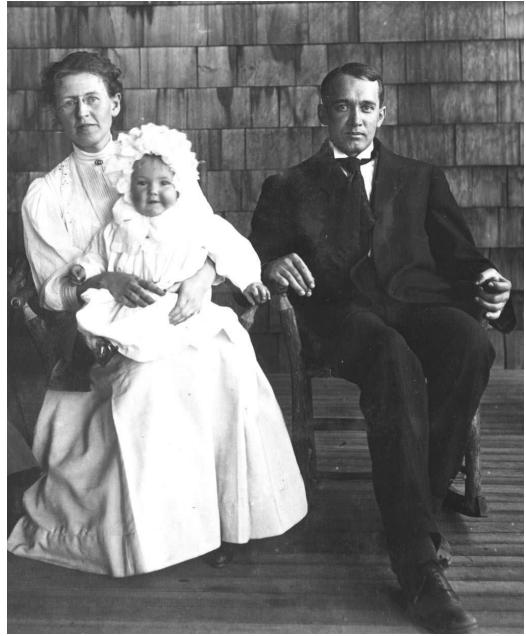}}
\caption{V.~M. Slipher, wife, daughter 1909 (courtesy LOA).}\label{tennfig04}
\end{figure}

   Slipher started buying rental properties and eventually bought a number of 
ranches. He owned and operated a furniture store at one time. He also took 
part in civic activities, serving as president of the school board when his 
children were in school and joining with others to found the Museum of 
Northern Arizona. He served as chairman of the board of Flagstaff's premier 
hotel, the Monte Vista. Founded by public subscription, including a major 
donation from author Zane Grey, it was built by the city in 1927. 
Meanwhile the Sliphers raised their family on Mars Hill (Figure \ref{tennfig04}).

   Despite these distractions, Slipher continued to be astronomically 
productive. It was during this period that he published observations of the 
spectra of both galactic and extragalactic nebulae and measured rotation 
speeds of planets. He made his only forays into solar astronomy, leading 
eclipse expeditions to Kansas in 1918 
\citep{Slipher1922ApJ....55...73S} and Baja California in 1923 to photograph the 
spectrum of the solar corona, and he continued the Lowell-inspired search for 
water, oxygen and chlorophyll on Mars \citep{Slipher1924PASP...36..261S}.
He also published his first investigations of 
the spectra of lightning, the aurora, and the night sky
\citep{Slipher1916LowOB...3....1S,Slipher1917LowOB...3...55S,Slipher1919ApJ....49..266S}.
   
\section{The Last Productive Years, 1926--33}

Slipher was appointed Director of the Lowell Observatory in 1926, after the 
final settling of the Lowell estate \citep{Smith1994}. Mrs. Lowell 
continued to receive some of the income until her death in 1954. 
Correspondence between V.~M. and trustee Roger Lowell Putnam shows that 
the Observatory was frequently in the red, and paychecks were not always issued on 
time. The trustee often helped with an extra check for \$500 or \$1000, and he 
even got his mother to pay the salary so that the observatory could have a 
secretary. The Observatory staff lost heavily when Flagstaff's only bank failed 
in 1932 \citep{Giclas1987Interview}. 

   During this period Slipher published his most extensive work on the night 
sky, zodiacal light, and the aurora, and he published additional spectroscopic 
observations of Venus, Mars, and comets. Every time C.~E. Kenneth Mees of 
Eastman Kodak came up with an emulsion that was sensitive a little farther 
out into the infrared Slipher used it to extend his planetary observations. One 
of his night sky spectrograms involved exposures totaling 147 hours! 
He also continued his observations of nebulae and studies of the 
interstellar medium. 

   His most famous work during this period was again an effort to carry on 
the work of his master. Lowell had spent years computing orbits (with much 
of the tedious calculation done by assistants, especially Elizabeth Williams) 
and attempting to make a prediction that would lead to the discovery of a 
ninth planet that would account for the perturbations of Neptune. In his highly 
mathematical 1915 book, \emph{Memoir on a Trans--Neptunian Planet}
\citep{Lowell1915}, he called it Planet X. He had employed as many as five 
computers in Flagstaff and Boston and had hired three successive ``Lawrence 
Fellows" to observe in Flagstaff searching for the planet, but he died without 
knowing whether it existed. In his book he had suggested two regions of the 
ecliptic, in opposite directions, where the planet might be found.

\begin{figure}
\center{\includegraphics[scale=0.7]{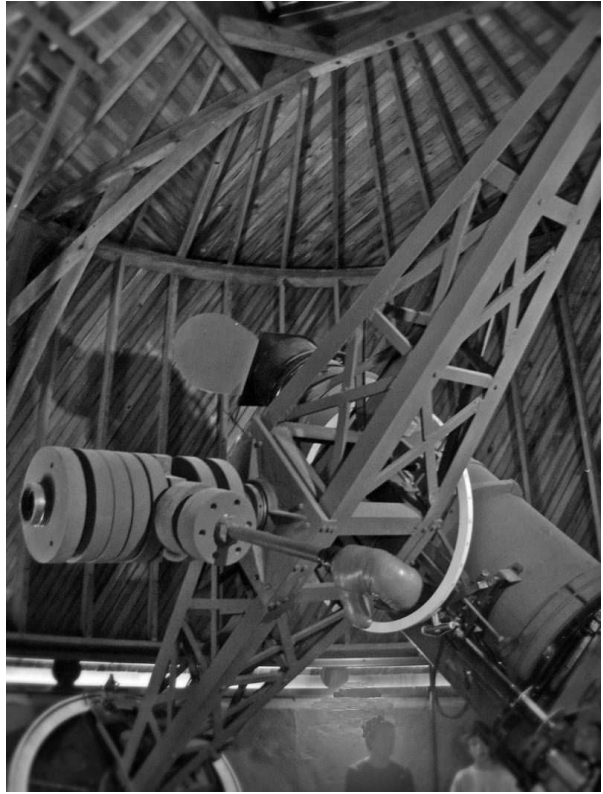}}
\caption{The Lawrence Lowell Telescope. (Courtesy Traci Lehman.)}\label{tennfig05}
\end{figure}

   In 1927 glass disks for a 13-inch refractor became available, and Slipher 
suggested to the trustee that they be purchased and made into a telescope to 
resume the search for Planet X. Trustee Guy Lowell personally purchased the 
disks and planned to have the telescope made, but he died later that year. At 
this point another member of the Lowell family, Percival's brother A. 
Lawrence Lowell, then president of Harvard, stepped in and funded the 
building of the telescope. It arrived in Flagstaff in 1929 and was erected on a 
mounting built by the observatory's longtime instrument maker, Stanley 
Sykes. Designed from the start for the planet search, the Lawrence Lowell 
telescope (Figure \ref{tennfig05}) produced highly defined star images over 14 x 17-inch 
plates. It could record 50,000 to 500,000 stars in a one-hour exposure. The 
plan was to photograph every field along the ecliptic, starting with the areas 
suggested by Percival Lowell, to repeat a few days later, and then to ``blink" the 
plates in order to find objects that moved.

\begin{figure}
\center{\includegraphics[scale=0.8]{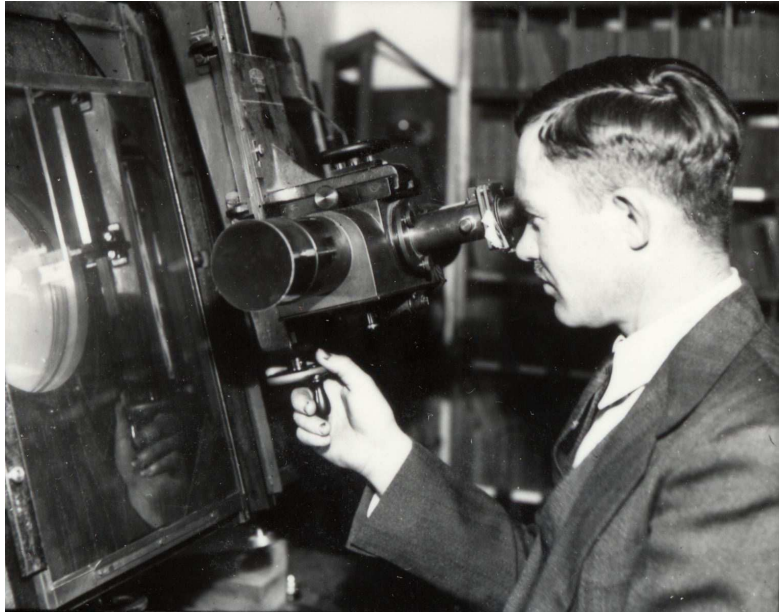}}
\caption{Clyde Tombaugh at the blink comparator (courtesy LOA).}\label{tennfig06}
\end{figure}

   Blinking the plates was incredibly tedious work. Slipher hired 23-year-old 
high school graduate Clyde Tombaugh (Figure \ref{tennfig06}) to do it, and the rest is 
history.\footnote{See, e.g., \cite{Hoyt1980QB701.H69.....H}}
 
   It is to Slipher's credit that Pluto is universally recognized as having been 
discovered by young Tombaugh. Had it been found at one of several other 
major observatories at the time, the director would have claimed credit. This is 
characteristic of Slipher's modesty.

   It is to Slipher's discredit that, knowing that the Lowell Observatory 
lacked the expertise to compute the orbit of the newly-discovered planet, he 
delayed announcing the positions so that his former teacher, John A. Miller of 
Sproul Observatory, could come to Flagstaff and lead the computation of the 
first orbit. Three days after the public announcement of the discovery (itself 
held until Lowell's birthday), a telegram to the Trustee signed ``Lowell Staff" 
reported \citep{LowellStaff1930}:
\begin{quote}
Impressed with vital importance to Observatory that our discovery 
announcement be followed soonest possible by best determined orbit our 
observations  can give because orbit will  demonstrate  much about nature 
and status of new wanderer that we telegraphed Professor Miller Director 
Sproul observatory asking him come Flagstaff and help us work best possible 
orbit. Miller experienced with  orbits loyal friend. Plans kept confidential.
\end{quote}

   Trustee Roger Putnam replied \citep{Putnam1930}, ``Frankly, I 
am very uncertain as to the ethics of when and what should be released, and 
will leave that to your judgment. I can't help feeling that having gotten the 
whole world stirred up, we have got to give them the information they want, 
but you know that sort of thing much better than I do." 

   Putnam was right, but Slipher delayed announcing more positions for four 
weeks while he and his colleague, C.~O. Lampland, frantically worked their 
slide rules under Miller's direction until they had an orbit. This infuriated 
many in the astronomical community, such as the ace orbit-computers at 
Berkeley, who could have produced an orbit much more quickly
\citep{Giclas1987Interview}. Slipher defended himself in a letter to the Trustee 
\citep{Slipher1930}:
\begin{quote}
We have been severely criticised for not giving out positions that others 
might comput [sic] the orbit, and this will no doubt not stop for a while yet. 
However, unpleasant as that has been it seemed our clear duty to make use of 
our materila [sic] for the orbit as it was more useful to us than it could be 
made to others without still more delay. Of course others could have done the 
orbit quicker than we did it, but we did it as carefully as possible. To have 
followed the other policy would have meant a considerable sacrifice to the 
Observatory.
\end{quote}

This is characteristic of Slipher's intense loyalty to Lowell Observatory and 
the memory of Percival Lowell.

\section{The Doldrums, 1934--54}

   The year 1933, when he turned 58, was essentially the last year that 
Slipher published his own original research. His five--page article on ``Spectra 
of the Night Sky, the Zodiacal Light, the Aurora, and the Cosmic Radiations 
of the Sky" appeared in the \emph{Transactions of the American Geophysical Union}
and was reprinted in the \emph{Journal of the Royal Astronomical Society of Canada}
\citep{Slipher1933JRASC..27..365S}. It reports on many years of work, 
including the use of a newly-designed spectrograph to photograph the spectra 
of five regions of the sky at once. He gave the George Darwin Lecture to the 
Royal Astronomical Society after accepting the RAS Gold Medal that same 
year. The lecture was on spectroscopic studies of the planets and summed up 
his work, mostly completed long before \citep{Slipher1933MNRAS..93..657S}.

\begin{figure}
\centering
\begin{tabular}{ccc}
\includegraphics[scale=0.8]{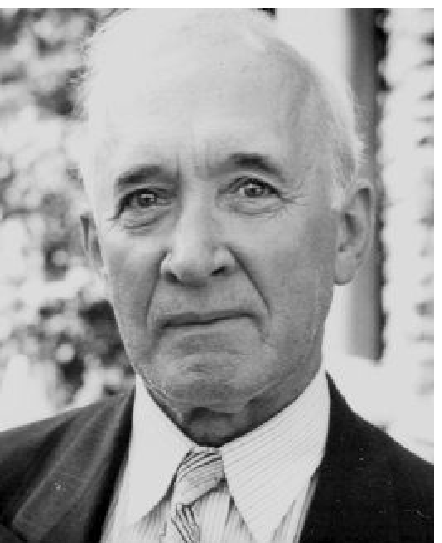}
\includegraphics[scale=0.8]{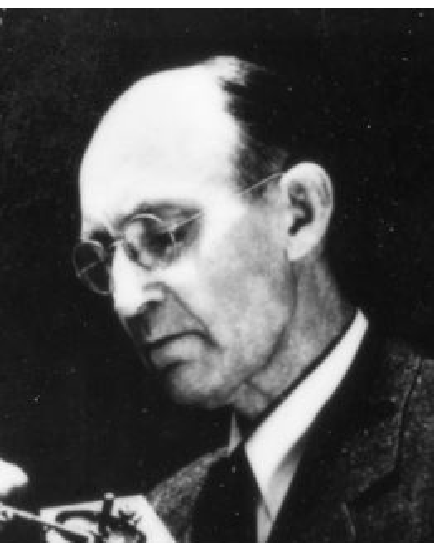} 
\includegraphics[scale=0.8]{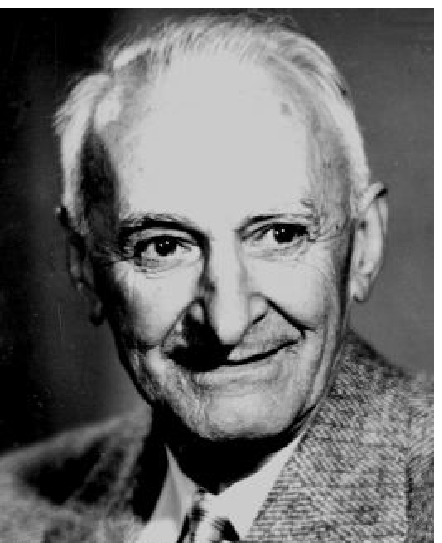} \\
\end{tabular}
\caption{V.~M. Slipher, C.~O. Lampland, and E.~C. Slipher in later years (courtesy LOA).}\label{tennfig07}
\end{figure}

   After that the Lowell Observatory slowly declined. For many years, three 
men -- V.~M., C.~O. Lampland, and V.~M.'s younger brother, E.~C. Slipher 
(Figure \ref{tennfig07}) -- dominated the observatory. Occasionally, a younger man, such as 
Henry Giclas in 1931, would be hired to a subordinate position, but the three 
senior astronomers jealously guarded the telescopes. All had been seriously 
wounded by the criticism from other astronomers, especially Lick 
Observatory directors W.~W. Campbell and W.~H. Wright, of work done at the 
Lowell Observatory. These eminent astronomers had developed an intense 
distaste for Percival Lowell and anything associated with him. Work coming 
from Lowell's observatory was automatically suspect.\footnote{Campbell did
come to respect Slipher in later years, but by then Slipher's habits were set.}
V.~M. had gotten into an exchange of criticisms with Campbell over his claim to have detected water 
on Mars in 1909, and after losing this battle he became even more reticent 
than he had been. He was very careful to check his work many times and to 
get repeated observations before going public. He devoted more and more 
time to his business affairs, and less to research. Meanwhile his brother, E.~C., 
spent much of his time on politics, and Lampland puttered around without 
completing anything. V.~M. published his last \emph{Observatory Report} in 1933. It 
covered the years 1930--1932. The next to appear from the Lowell Observatory 
was for the years 1952--1954. Although signed by V.~M., it was probably 
written by Albert G. Wilson, who was assistant director at the time.
The most significant papers with Slipher's name on them after 1933 were of a 
totally different character from his other work. There were six of them, 
published in  \emph{Nature} and the \emph{Physical Review}, and they contained 
astrophysical observations and theory far beyond Slipher's
abilities.\footnote{See \cite{Adel1934PhRv...46Q.240A,
Adel1934PhRv...46R.240A,Adel1934Natur.134..148A,Adel1934PhRv...46..902A,Adel1935PhRv...47..580A,
Adel1935PhRv...47..651A}.} They were written by Arthur 
Adel, who had been hired in 1933 by the trustee over the opposition of the 
senior astronomers. Adel was to work at his alma mater, the University of 
Michigan, and do infrared studies that would relate Slipher's spectra to 
conditions on the planets. Adel built a 22.5-m long high pressure cell and put 
up to 40 atmospheres of carbon dioxide in it. Later he filled his tube with 
ammonia and methane. He was able to duplicate some of the spectra that Mt. 
Wilson astronomers had observed in Venus in 1932 and that V.~M. had 
observed in Jupiter many years earlier. He did this entirely by himself in Ann 
Arbor for \$1000 per year, which even in 1933 was not much.

   Adel used Slipher's published data but got nothing new from Slipher. 
Nevertheless, he put Slipher's name on the papers as co-author. In 1987 Adel 
told Robert Smith in an oral history interview \citep{Adel1987interview},
\begin{quote}
I had to do that, and neither he nor Lampland nor E.~C. Slipher, none of them 
really knew what I was doing, had a real understanding of it. ... They didn't 
know anything about infrared spectroscopy. They didn't know anything 
about spectroscopy. They really didn't know anything about this work that I 
was doing, or the work I did in Ann Arbor. 
\end{quote}

   When Adel was appointed to a lowly position in Flagstaff by the trustee, 
V.~M. treated him very badly. And when Adel showed that the carbon dioxide 
bands in the spectrum of Venus could be photographed with the 24-inch 
refractor and thus could have been discovered by Slipher before they were 
found by Adams and Dunham at Mt. Wilson \citep{Adams1932PASP...44..243A},
he was barred from all the telescopes \citep{Adel1987interview}.

   According to Henry Giclas \citep{Giclas1987Interview,Giclas1990},
Slipher resisted applying for grants and couldn't be 
bothered with the complications of payroll, social security, etc. Giclas was 
appointed executive secretary in 1953 and took over all the business affairs. 
Trustee Roger Putnam pushed for grants, and the first, from the Weather 
Bureau, was obtained in 1948 and included funds to measure the variation in 
the solar constant as well as meteorology of planetary atmospheres. Later this 
project was taken over by the Air Force. The appointment of Harold Johnson 
in July 1948, initially to work on the Weather Bureau project, was a turning 
point. Although very difficult to get along with and constantly complaining, 
he was a competent, energetic young scientist, and he accounted for nearly all 
of the Observatory's publications in the early 1950s. He quit and went to 
Yerkes after one year, but was hired back in August 1952 by the trustee over 
V.~M.'s objections.

   Slipher's last scientific publication was a brief abstract in 1939 
announcing that he had re-observed Hubble's variable nebula, NGC 2261, and 
found that its spectrum had not changed since his observations of 1916--17 
\citep{Slipher1939PASP...51..115S}. He also wrote an occasional letter 
asserting his priority on something done long before. 
 
\begin{figure}
\center{\includegraphics[scale=0.5]{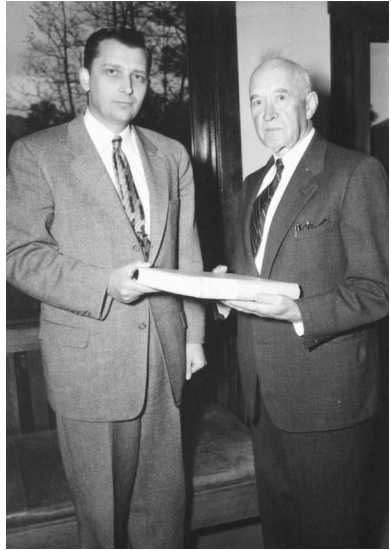}}
\caption{Albert G. Wilson and V.~M. Slipher at Slipher's retirement party, 1955 (courtesy LOA).}\label{tennfig08}
\end{figure}

\section{Retirement, 1954--69}

After Lampland died in December 1951, trustee Roger Putnam finally stepped 
in to make some changes. On the advice of John Duncan he selected
33-year-old Albert G. Wilson, who had been directing the National Geographic 
Palomar Sky Survey since completing his Ph.D. in mathematics at Caltech. It 
seems that the primary criterion for the appointment was that Wilson was 
acceptable to the Slipher brothers.

   Wilson came as assistant director in 1953 and took over as director on 
V.~M.'s 79th birthday, 11 November 1954, when the old man finally retired 
(Figure \ref{tennfig08}). Wilson's directorship was short and unhappy. After a rebellion 
from the younger astronomers, especially Harold Johnson and Henry Giclas, 
and the breakup of his marriage, he left in January 1957 and returned to 
California and a career in industry \citep{Tenn2007JAHH...10...65T}. 

   Slipher remained in Flagstaff although he moved off Mars Hill into one of 
his houses. His wife, Emma, died in 1961. Frances Wilson, the ex-wife of 
Slipher's successor, returned to Flagstaff and became Slipher's ``private 
secretary and companion" according to Henry Giclas \citep{Giclas1990}.
V.~M. Slipher died 8 November 1969, three days before he would have 
turned 94.

   His will \citep{Slipher1967} stated that ``During the latter years of my lifetime,\newline
FRANCES M. WILSON has devoted herself to my 
business affairs, and it is my desire from my Estate to make provision for 
her." He left her \$3000 per year for life, and he made her executrix of his 
estate. Aside from the endowment to support her he left his wealth to the V.~M. 
Slipher Trust with a bank as trustee and Arthur Adel
\footnote{The appointment of Adel to select the scholarship recipients and
Adel's service as a pallbearer at Slipher's funeral are remarkable,
considering that Adel was vehemently critical of how Slipher had
mistreated him in the 1930s \citep{Adel1987interview}.
When I met Adel at the Lowell Observatory centennial in 1994 he was
still resentful. Adel spent most of his life in Flagstaff, where he had a
highly successful career at Northern Arizona University after World War II.}
as Advisory Trustee. A 
portion of the income was to provide scholarships for worthy students 
pursuing scientific studies at Arizona's three public universities. After that 
50\% of income went to the National Academy of Sciences for Astronomy. 
There was a great deal of property, including ranches and cattle.

\section{Conclusion}

   Although he received prestigious awards in his lifetime, including the 
1935 Bruce Medal of the Astronomical Society of the Pacific and the three 
mentioned in the obituary below, Slipher is probably underrated today. I gave 
a talk \citep{Tenn2005AAS...207.5804T} at a meeting of the Historical 
Astronomy Division of the American Astronomical Society in 2006 titled 
``Why Does V.~M. Slipher Get So Little Respect?" My current conclusion is 
that the most important reasons are

\begin{enumerate}
\item He needed Lowell to guide him, and Lowell's early death left him 
unprepared to face the future. Although a skilled spectroscopist, he lacked the 
imagination to innovate.

\item The early criticism of Lowell and everyone around him made Slipher and 
his colleagues super-cautious about making any claims. They hesitated to 
publish until they were absolutely certain they were right. Fortunately, this 
happened a few times with V.~M. His brother, E.~C., would never have 
published had the trustee not forced him to. The result was a fine atlas of 
photographs of Mars. The third member of the Lowell staff, C.~O. Lampland, 
did pioneering work in radiometry (infrared photometry), but hardly ever 
published. All three stayed too long, at least in part because there were no 
pensions until they were introduced by V.~M.'s successor, Wilson, in the 1950s.

\item The Lowell Observatory's poverty from the death of its founder in 1916 
until after Slipher's retirement precluded buying modern equipment that could 
compete with the Mt. Wilson and Lick Observatories in California and also 
led to Slipher turning much of his attention toward improving his personal 
finances. 

\item He was not properly credited by Hubble for the Doppler shifts of galaxies 
that Hubble used so successfully in his key
1929 paper \citep{Hubble1929PNAS...15..168H}. All of the credit went to Hubble and Milton 
Humason. (Hubble did credit Slipher in later papers, starting in 1931.) 

\end{enumerate}

Slipher's brief obituary
in \emph{Physics Today} \citep{Anonymous1970PhT....23b.101.},
which mentions only the discovery of Pluto 
among his accomplishments, makes this clear. It reads, in its entirety:
\begin{quote}
Vesto M. Slipher, director of the Lowell Observatory until 1952 [sic], died 8 Nov. 
at 93. Slipher had been at the observatory since 1901 and became director in 
1926. He supervised work that led to the discovery in 1930 of Pluto. Among 
the honors received by Slipher were the Lalande Prize and gold medal of the 
Paris Academy of Sciences (1919), the Draper Medal of the National 
Academy of Sciences (1932) and the Royal Astronomical Society gold medal (1932).
\end{quote}

\acknowledgements

I thank Lauren Amundson, Antoinette Beiser, and Martin\newline
Hecht of the Lowell 
Observatory Archives for documents and images and Traci Lehman for one 
image. I benefited from helpful discussions with Arthur Adel (1994), Frank 
Edmondson (2005), Henry L. Giclas (2007), and Albert G. Wilson (2005, 
2012). I also appreciate David DeVorkin's helpful comments on this article. 
This research has made extensive use of NASA's Astrophysics Data System.

\bibliography{tenn}

\end{document}